\documentclass[oneside,a4paper,11pt,shownumbers]{article}
\usepackage{latexsym}
\usepackage{euscript}
\usepackage{epsfig,amsmath,amssymb}

\def\be{\begin{equation}}
\def\ee{\end{equation}}
\def\bea{\begin{eqnarray}}
\def\eea{\end{eqnarray}}

\long\def\symbolfootnote[#1]#2{\begingroup%
\def\thefootnote{\fnsymbol{footnote}}\footnote[#1]{#2}\endgroup} 

 \large\normalsize

\begin{document}

\title{\LARGE \bf Exact String-Like Solutions in Conformal Gravity}
  \author{
  \large  Y. Verbin$^a$ $\:${\em and}$\:$
 Y. Brihaye$^b$ \thanks{Electronic addresses: verbin@openu.ac.il; Yves.Brihaye@umons.ac.be } }
 \date{ }
   \maketitle
   \centerline{$^b$ \em Department of Natural Sciences, The Open University
   of Israel,}
   \centerline{\em Raanana 43107, Israel}
     \vskip 0.4cm
      \centerline{$^b$ \em Physique Th\'eorique et Math\'ematiques, Universit\'e de Mons,}
   \centerline{\em Place du Parc, B-7000  Mons, Belgique}

\maketitle
\thispagestyle{empty}   

\begin{abstract}
The cylindrically-symmetric static vacuum equations of Conformal Gravity are solved for the case of additional
boost symmetry along the axis. We present the complete family of solutions which describe the exterior 
gravitational field of line sources in Conformal Gravity. We also analyze the null geodesics in these spaces.\\


\end{abstract}

\maketitle
\medskip \medskip

\section{Introduction}\label{Introduction}
\setcounter{equation}{0}

Conformal Gravity (CG) \cite{Mannheim2006} was proposed as a possible alternative to Einstein gravity (``GR''), which may 
supply the proper framework for a solution to some of the most annoying problems of theoretical physics like those of the 
cosmological constant, the dark matter and the dark energy.

The gravitational field  in CG is still minimally coupled to matter, but the dynamical basis is different: it is obtained 
by  replacing the Einstein-Hilbert action with the Weyl action based on the Weyl (or {\it conformal}) tensor 
 $C_{\kappa\lambda\mu\nu}$ 
defined as the totally traceless part of the Riemann tensor:
\begin{eqnarray}
C_{\kappa\lambda\mu\nu}=R_{\kappa\lambda\mu\nu}-
\frac{1}{2}(g_{\kappa\mu}R_{\lambda\nu}-g_{\kappa\nu}R_{\lambda\mu}+
g_{\lambda\nu}R_{\kappa\mu}-g_{\lambda\mu}R_{\kappa\nu})+
\frac{R}{6}(g_{\kappa\mu}g_{\lambda\nu}-g_{\kappa\nu}g_{\lambda\mu})
\label{WeylTensor},
\end{eqnarray}
so the gravitational Lagrangian is
\begin{equation}
{\cal L}_{g}= -\frac{1}{2\alpha}C_{\kappa\lambda\mu\nu}C^{\kappa\lambda\mu\nu} 
\label{GravL}
\end{equation}
where $\alpha$ is a dimensionless parameter. The gravitational field equations take the following form:
\begin{equation}
W_{\mu\nu} =  \frac{\alpha}{2} T_{\mu\nu} 
\label{GravFieldEq}
\end{equation}
where $T_{\mu\nu}$ is the energy-momentum tensor and $W_{\mu\nu} $ is
the Bach tensor given by
\be
W^{\mu\nu}= R_{\kappa\lambda}C^{\kappa\mu\lambda\nu}-2\nabla_\kappa\nabla_\lambda C^{\kappa\mu\lambda\nu}
\label{BachTensorInTermsOfC},
\ee
or in terms of the Riemann and Ricci components by:
\begin{eqnarray}
W_{\mu\nu}=\frac{1}{3}\nabla_\mu\nabla_\nu R-\nabla_\lambda\nabla^\lambda R_{\mu\nu}
+\frac{1}{6} (R^2+\nabla_\lambda\nabla^\lambda R-3R_{\kappa\lambda}R^{\kappa\lambda})g_{\mu\nu}+
2R^{\kappa\lambda}R_{\mu\kappa\nu\lambda}-\frac{2}{3}RR_{\mu\nu}
\label{BachTensor}.
\end{eqnarray}

It was suggested (see \cite{Mannheim2006} and references therein) that while CG agrees with Newtonian gravity in 
Solar System scales, it further produces a linearly growing potential that could explain galactic rotation curves 
without invoking dark matter. It was further argued that accelerating cosmological solutions of CG describe naturally the 
accelerated expansion of the universe thus removing the need for dark energy.

On the other hand, CG has been criticized from several aspects both phenomenological and formal. 
Arguments in favor of the need of dark matter come from observations of the unusual object called
``bullet cluster'' \cite{CloweEtAl2006,BradacEtAl2006} whose dynamics seems very difficult to understand without
assuming a weekly interacting dark component.

More specifically, several authors claim that predictions in the weak field limit of CG disagree 
with solar system observations \cite{Flanagan}, yield wrong light deflection \cite{EderyPar1998} and
that the exterior solutions cannot be matched to any source with a ``reasonable'' 
mass distribution \cite{PerlickXu}. Other authors find evidence for tachyons or ghosts 
\cite{BarabashSht1999} or raise the fact that only  null geodesics are physically meaningful in this 
theory since the ``standard'' point particle Lagrangian is not conformally-invariant \cite{WoodMoreau,BrihayeVerbinSph}.

Counter arguments to some of these objections were also published \cite{EderyEtAl2001,Mannheim2007}, as well as
possible ways \cite{BrihayeVerbinSph} out of some of the difficulties and the matter is, to 
our view, still waiting for a consensus. 

It is therefore very much required to investigate further the predictions and consequences of CG in its purely tensorial
formulation as well as in its scalar tensor extension  as much as possible. 

In this work we concentrate on cylindrically-symmetric static vacuum solutions with the aim of clarifying further the 
properties of string-like solutions in CG \cite{BrihayeVerbinCyl}. Cosmic strings \cite{VilSh} 
are a typical outcome in any field theory which describes matter in the very early universe, thus serving very 
well the purpose of testing the implications of CG.  The main result reported here is the full family of the 
 static cylindrically-symmetric vacuum solutions of CG which represent the external gravitational field of localized line
 sources.
 
\section{Cylindrically-Symmetric Equations} \label{CylSymEqs}
\setcounter{equation}{0}

The general static cylindrically symmetric line-element has the form:
\begin{equation}
ds^2= B^2 (r)dt^2 - M^2 (r)dr^2 - L^2 (r)d\varphi^2 - K^2 (r)dz^2
\label{lineel}
\end{equation}

 In order to find solutions for this system, we have to fix the arbitrariness of the radial 
coordinate and the arbitrary rescaling of the metric due to the conformal symmetry. 
We will also limit our solutions to those exhibiting boost symmetry along the string direction 
($z$), as for ordinary cosmic strings, i.e. $K(r)=B(r)$. This leaves one independent metric component. 
So only one of the gravitational field equations (\ref{GravFieldEq}) (with $T^{\mu}_{\nu}=0$) has to be solved. 
We may therefore choose to solve the lower order $rr$ equation, i.e. $W_r^r  =0$.

Several special solutions are already known in explicit form. First of all, all the thin string (line source) 
solutions of GR with either a vanishing or non-vanishing cosmological constant \cite{linet} which solve 
for $r>0$
\be
R^{\mu}_{\nu}= \frac{\Lambda}{4}\delta^{\mu}_{\nu}
\label{AdSEq} 
\ee
satisfy also the CG vacuum equations $W_{\mu\nu}=0$. This is obvious by direct substitution of (\ref{AdSEq}) 
in Eq. (\ref{BachTensor}). 

One special member of the $\Lambda<0$ family is  the  AdS soliton \cite{HorMy1999} (see also \cite{bbh,Bonnor,tekin}) 
which is a cylindrically-symmetric regular solution of the same equation (\ref{AdSEq}) and is therefore distinct from 
AdS (anti-de Sitter) space.  More recently, we have discovered two families of very simple exact 
solutions \cite{BrihayeVerbinCyl} during a mainly numerical 
study of cylindrical solutions of the Abelian Higgs model coupled to CG. However, we were unable to integrate the equations
analytically at the time.

Here we present a reduction of the vacuum equations for static cylindrically-symmetric solutions to a single first
order non-linear equation which may be solved by a straightforward quadrature. 

We have therefore first to complete the gauge fixing. Since we would like it to be consistent with 
the symmetric vacuum solutions [i.e. (A)dS spaces] and with the AdS soliton, it should respect the asymptotic
condition 
\begin{equation}
R^{\mu}_{\nu}\rightarrow \frac{\kappa}{4}\delta^{\mu}_{\nu} \;\;\;\; \text{for} \;\; r\rightarrow \infty
 \label{AsymptAdS}
\end{equation}
where $\kappa$ is a real parameter. Note however, that a constant Ricci scalar is not a ``gauge invariant'' concept 
in CG; it is only a matter of convenience which can be obtained by a proper gauge choice. The corresponding
invariant condition is that the Weyl tensor will vanish asymptotically, that is spacetime  becomes  
conformally flat asymptotically. 
These restrictions will simplify considerably the very cumbersome expressions 
of the components of the Bach tensor and will enable a clear physical picture.

The simplest gauge choice is $B(r)=K(r)=M(r)=1$ with $L(r)$ as a single metric component. This choice
may be obtained directly from Eq. (\ref{lineel}) by a suitable conformal transformation combined with a redefinition 
of the radial coordinate. In this gauge $R^0_0=R^z_z = 0$ \footnote{see the discussion about the metric 
$\hat{g}_{\mu\nu}=diag(1,-1,-H^2,-1)$ at the end of this section} so it cannot contain the well-known solutions with a 
cosmological constant of Eq. (\ref{AdSEq}). The cylindrical version of the ``Mannheim gauge''\cite{Mannheim2006}  
$
 M(r)=1/B(r) \; , L(r) = r 
$
with $K(r) = B(r)$, turns out to be complicated and the simplifications from spherical symmetry are not 
observed.

An alternative parametrization of the metric tensor which solves these difficulties is the 
following gauge choice
\be
\label{gauge}
      M = 1 \ \ , \ \ L=\frac{dB}{dr} = B' \ \ , \ \ K(r) = B(r)
\ee
where we also resort to dimensionless coordinates.  This metric is equivalent to that of the ``Mannheim gauge'',
but has the advantage that it leads to autonomous equations which we can solve by quadrature.

Note however, that this gauge excludes the ``flat'' $\Lambda=0$ solutions. Conformally flat solutions are 
of course still allowed.

In this gauge the components of Ricci tensor and scalar take the form 
\be
 R_0^0 = R_z^z = -\left(\frac{B'}{B}\right)^2 - 2\frac{B''}{B}  \  , \ 
 R_r^r = R_\varphi^\varphi = -\frac{B'''}{B'} - 2\frac{B''}{B} \ , \ 
 R = -2 \left[\left(\frac{B'}{B}\right)^2 + 4 \frac{B''}{B} + \frac{B'''}{B'}\right]
 \label{Ricci-BG}
\ee
while the rest vanish. 

Already at this stage we can obtain easily the constant Ricci solutions of Eq. (\ref{AdSEq}). 
It is enough to solve the (00) equation which is readily integrated to give 
\be
\frac{1}{2}(B')^2 + \frac{\Lambda}{24}B^2 - \frac{b}{B} =0
 \label{ConstRicci-BG}
\ee 
where $b$ is the integration constant. This is a trivial ``mechanical'' equation whose solutions are easily obtained by
inspection. We will discuss these solutions within the more general framework in the next section.

The non-zero components of the Weyl tensor are all proportional to a single quantity
namely
\be
C^{0r}_{\ \ 0r}=C^{0\varphi}_{\ \ \ 0\varphi}=C^{rz}_{\ \ \ rz}=C^{\varphi z}_{\ \ \ \varphi z}={\cal C}/6 \ \ , 
\ \ C^{r\varphi}_{\ \ \ r\varphi}=C^{0z}_{\ \ \ 0z}=-{\cal C}/3 
 \label{WeylTensorComp}
\ee
where   
\be
{\cal C} = \frac{B'''}{B'}-2\frac{B''}{B}+\left(\frac{B'}{B}\right)^2
 \label{WeylComp},
\ee
and its square is given simply by $C_{\kappa\lambda\mu\nu}C^{\kappa\lambda\mu\nu}=4{\cal C}^2/3$.

the non-vanishing components of the Bach tensor are 
\begin{eqnarray} \nonumber
W_0^0  =W_z^z  =
\frac{B^{(5)}}{3 B'}+\frac{2 B^{(4)}}{3 B}-
\frac{B^{(4)} B''}{3 B'^2}+\frac{B''' B''^2}{3 B'^3}+\frac{2 B''' B''}{3 B B'}-\frac{2 B''' B'}{3 B^2}\\
-\frac{2 B'''^2}{3 B'^2}-\frac{4 B''^2}{3 B^2}+\frac{4 B'^2 B''}{3 B^3}-
\frac{B'^4}{3 B^4}
\label{Bach00}
\end{eqnarray}

\begin{eqnarray} \nonumber
W_r^r  =\frac{2 B^{(4)}}{3 B}-
   \frac{2 B^{(4)} B''}{3 B'^2}+\frac{2 B''' B''^2}{3 B'^3}-\frac{2 B'''B''}{B B'} +\frac{2 B''' B'}{3B^2}\\
  +\frac{B'''^2}{3B'^2}
   +\frac{4 B''^2}{3 B^2}-\frac{4 B'^2 B''}{3B^3}+
\frac{B'^4}{3B^4}
\label{Bachrr}
\end{eqnarray}
while the fourth one, $W^\varphi_\varphi$ can be obtained immediately from the identity $W^\mu_\mu =0$.

Using an exponential transformation $B=e^\beta$ simplifies the expressions a little. $W_r^r$ and the ``Weyl tensor 
quantity'' ${\cal C}$ which will be used in the following, become:
\begin{eqnarray} 
W_r^r  = -\beta ''^2 +\frac{\beta^{(3)2}}{3 \beta'^2}+\frac{2\beta ^{(3)} \beta''^2}{3\beta '^3}-
\frac{4 \beta ^{(3)} \beta ''}{3\beta '}-\frac{2\beta ^{(4)}\beta ''}{3\beta '^2} \ \ , \ \ \ \ \ \ 
{\cal C} = \frac{\beta'''}{\beta'}+\beta''
\label{Bachrrbeta}
\end{eqnarray}

An easy way to obtain $W_\mu^\nu$ is to start with the line element 
$ds^2=B^2 [dt^2-d\varrho^2-H^2 d\varphi^2-dz^2]$ whose Bach tensor is given by $W_\mu^\nu=\hat{W}_\mu^\nu/B^4$
where $\hat{W}_\mu^\nu$ is calculated from the metric $\hat{g}_{\mu\nu}=diag(1,-1,-H^2,-1)$. The components
$\hat{W}_\mu^\nu$ are relatively easy to obtain since $\hat{g}_{\mu\nu}$ has only one non-vanishing independent 
Riemann component: $\hat{R}_{\varrho\varphi \varrho\varphi}=-H d^2 H/d\varrho^2$. We thus get
\begin{eqnarray}\nonumber
 \hat{W}_0^0 = \hat{W}_z^z = \frac{1}{3}\left[\frac{H''''}{H} - \frac{H'H'''}{H^2} -
 2\left(\frac{H''}{H} \right)^2  + \frac{H'^2H''}{H^3} \right] \  , \ \\
 \hat{W}_\varrho^\varrho = \frac{1}{3}\left[ - 2\frac{H'H'''}{H^2} +
 \left(\frac{H''}{H} \right)^2  + 2\frac{H'^2H''}{H^3} \right] , \ \ \ \ \ \ '=\frac{d}{d\varrho}  
  \label{Bach-Hat}
\end{eqnarray}
while $\hat{W}^\varphi_\varphi$ can be obtained from $W^\mu_\mu =0$. A further coordinate transformation 
$Bd\varrho = dr$ together with $H=d\beta/dr$ yields after lengthy calculations the above components of the Bach tensor
of Eqs. (\ref{Bach00})-(\ref{Bachrr}). As a check we have verified that these components satisfy a covariant conservation 
law, $\nabla_{\mu}W^\mu_\nu =0$ which in the present case reduces to the single equation
\be
B B'(W_r^r)'+2(B B''+B'^2)W_r^r+2(B B''-B'^2)W_0^0=0
\label{Wconservation}.
\ee
We also calculated $W^\varphi_\varphi$ explicitly and verified that indeed $W^\mu_\mu =0$.

\section{String-Like Solutions}\label{String-Like Solutions}
\setcounter{equation}{0}

Solving the equations becomes possible thanks to the following identity
\begin{eqnarray}
W_r^r = -\frac{2}{3}\frac{e^{-3 \beta /2}}{\beta '} \left[\beta ''\left( e^{3 \beta /2}
   \left(\beta ''+\frac{\beta'''}{\beta '}\right)\right)'-
   \frac{1}{2} \beta''' e^{3 \beta /2}
   \left(\beta ''+\frac{\beta'''}{\beta '} \right) \right]
 \label{WrrIdentity}
\end{eqnarray}
which may be written as a total derivative in the following form:
\begin{eqnarray}
W_r^r = -\frac{2\epsilon}{3}\frac{\left(\epsilon\beta ''\right)^{3/2} e^{-3 \beta /2}}{\beta '} \left[\frac{e^{3 \beta /2}
  }{\sqrt{\epsilon\beta ''}} \left(\beta ''+\frac{\beta'''}{\beta '}\right)\right]'
  =\frac{4}{3}\frac{\left(\epsilon\beta ''\right)^{3/2} e^{-3 \beta /2}}{\beta '}
  \left[ \frac{\left(\sqrt{\epsilon e^{\beta}\beta ''}\right)'}{\left(e^{- \beta}\right)'}  \right]'
 \label{WrrIdentity2}
\end{eqnarray}
where $\epsilon=\pm 1$ according to whether $\beta ''$ is positive or negative. The motivation to look for such 
identities comes from the fact that the components of Bach tensor
may be expressed in terms of the Weyl tensor as in Eq. (\ref{BachTensorInTermsOfC}). Although it may seem
simpler to get solutions in other gauges like the Mannheim gauge or in terms of the metric component $H(r)$ defined
above, we were able to integrate the equations in terms of $B(r)$ (or $\beta(r)$) only.

Solving now the equation $W_r^r =0$ is straightforward and quite simple. We will use both $\beta$ and $B$ alternatively
according to convenience. The first kind of solutions satisfies $\beta ''=0$ which is solved by
\be
B(r)=B_0 e^{kr}
\label{SolExp}
\ee 
where $k$ may be either positive or negative. 

The only other possibility is that the ratio $(\sqrt{\epsilon e^{\beta}\beta ''})'/(e^{- \beta})'$ 
 in the second factor of (\ref{WrrIdentity2}) is a constant, say $c$, so we obtain after two integrations the following
  ``mechanical'' equation
\be
\frac{1}{2}(\beta ')^2+\frac{\epsilon}{3c}(a+ce^{-\beta })^3 = E \,\,\, ,\,\,\, c\ne 0 \,\,\,\,\,\, ;\;\;\;\;\;\; 
\frac{1}{2}(\beta ')^2+\epsilon a^2 e^{-\beta } = E\,\,\, ,\,\,\,c=0 
\label{FirstOrderMechbeta}
\ee 
where $a$ and $E$ are integration constants. Note that the special case $c=0$ needs special care. 
A special family of solutions of this case is the above (\ref{SolExp}). Others will be considered later.
Actually, it is easy to see from Eqs (\ref{Bachrrbeta}) and (\ref{WrrIdentity2}) that $c=0$ corresponds to 
conformally flat solutions with vanishing Weyl tensor.

\begin{figure}[!t]
\centering
\leavevmode\epsfxsize=10.0cm
\epsfbox{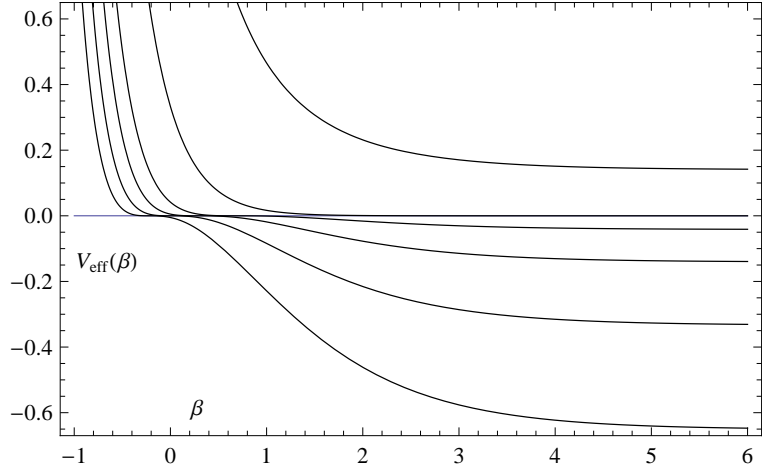}\\
\caption{\label{FigVeff} \small{The effective potential $V_{eff}(\beta)$ for $\epsilon=1$, $c=1$ and 
$a=-1.25,\, -1, \,-0.75, \,-0.5, \,0, \,0.75$. In order to identify the curves, note that the asymptotic value 
of $V_{eff}(\beta)$ increases with $a/c$ for $\epsilon=1$.}}
\end{figure} 

 We therefore
define an ``effective potential'' $V_{eff}(\beta)$ which satisfies $(\beta ')^2 /2+ V_{eff}(\beta)=E $:
\begin{eqnarray}
V_{eff}(\beta)=
 \begin{cases}
 \frac{\epsilon}{3c}(a+ce^{-\beta })^3 & \,\,\, ,\,\,\, c\ne 0 \\
 \epsilon a^2 e^{-\beta } & \,\,\, ,\,\,\, c= 0 \, .
\end{cases}
 \label{EffPotbeta}
 \end{eqnarray}
Fig. \ref{FigVeff} presents the general behavior of $V_{eff}(\beta)$ for $\epsilon=+1$. The curves for $\epsilon=-1$ 
are just the same taken ``upside-down''.
In terms of the metric function $B$ the equation becomes $(B ')^2 /2+ V_{eff}(B)=0 $ where now 
\begin{eqnarray}
V_{eff}(B)=
 \begin{cases}
 (\frac{\epsilon a^3}{3c}-E)B^2+\epsilon a^2 B +\frac{\epsilon c^2}{3B} +\epsilon ac & \,\,\, ,\,\,\, c\ne 0 \\
 \epsilon a^2 B-EB^2 & \,\,\, ,\,\,\, c= 0 \, .
\end{cases}
 \label{EffPotB}
 \end{eqnarray}

The solutions may be expressed directly in terms of the metric function $B$, but part of the presentation is clearer in 
terms of $\beta$. So we will use both forms of solutions
\be
r(B)=\int dB/\sqrt{-2V_{eff}(B)}\,\,\,\,\,\,;\,\,\, r(\beta)=\int d\beta/\sqrt{2\left(E-V_{eff}(\beta)\right)}
\label{r(B)}
\ee 
where the integration limits are determined by the boundary conditions.
 
By inspection of the potential function one can conclude that there are 3 distinct types of solutions: 
one for $\epsilon = +1$ and two for $\epsilon = -1$. It is easy to see that 
for $\epsilon = +1$ there are solutions only for values of $E$ obeying $E>a^3/3c$ (or $E>0$ for $c=0$). The effective 
potential $V_{eff}(\beta)$ decreases monotonically with $\beta$ and the solutions have a finite minimal value unless 
$c=0$. On the other hand, for $\epsilon = -1$, $E$ can have any real value. This case splits therefore according to whether 
$E$ is above or below the maximal value of the effective potential $V_{eff}(\beta)$ which is $-a^3/3c$ (or 0). 
If $E<-a^3/3c$, $\beta$ or $B$ is bounded from above. If $E\ge-a^3/3c$, the solutions can have the whole range 
$0\le B<\infty$.  For future use we define the ``energy excess'' parameter $\zeta$ by $3E/c^2=\zeta^3+\epsilon(a/c)^3$.

It is possible to express the solutions in terms of hyperelliptic integrals, but since the effective potential is monotonic,
the possible behaviors of the solutions are limited so it is enough to introduce the following 
two kinds of real functions:
\be
\Upsilon^{(+)}_{1}(\alpha,u)=\int_{1}^{u}\frac{dt}{t\sqrt{(\alpha+1)^3-(\alpha+1/t)^3}} 
\,\,\,\,\,\, ,\,\,\,  u \ge 1
\label{Ypsilon1pl}
\ee 
\be
\Upsilon^{(-)}_{1}(\alpha,u)=\int_{u}^{1}\frac{dt}{t\sqrt{(\alpha+1/t)^3-(\alpha+1)^3}}
\,\,\,\,\,\, ,\,\,\, 0 \le u \le 1
\label{Ypsilon1min}
\ee 
and
\be
\Upsilon^{(+)}_{2}(\gamma,u)=\int_{0}^{u}\frac{dt}{t\sqrt{(\gamma+1/t)^3-\gamma^3+1}} 
\,\,\,\,\,\, ,\,\,\,  u \ge 0
\label{Ypsilon2pl}
\ee 
\be
\Upsilon^{(-)}_{2}(\gamma,u)=\int_{0}^{u}\frac{dt}{t\sqrt{(\gamma+1/t)^3-\gamma^3-1}}
\,\,\,\,\,\, ,\,\,\, 0 \le u \le u_{max}
\label{Ypsilon2min}
\ee
where $u_{max}$ is the solution of $(\gamma+1/u)^3=\gamma^3+1$, that is:
\be
u_{max}=\frac{1}{\sqrt[3]{1+\gamma^3}-\gamma}
\label{umax}
\ee
with the cube root $\sqrt[3]x$ means $-|x|^{1/3}$ for $x<0$.
\begin{figure}[!t]
\centering
\leavevmode\epsfxsize=10.0cm
 \includegraphics[height=.28\textheight,width=.48\textwidth]{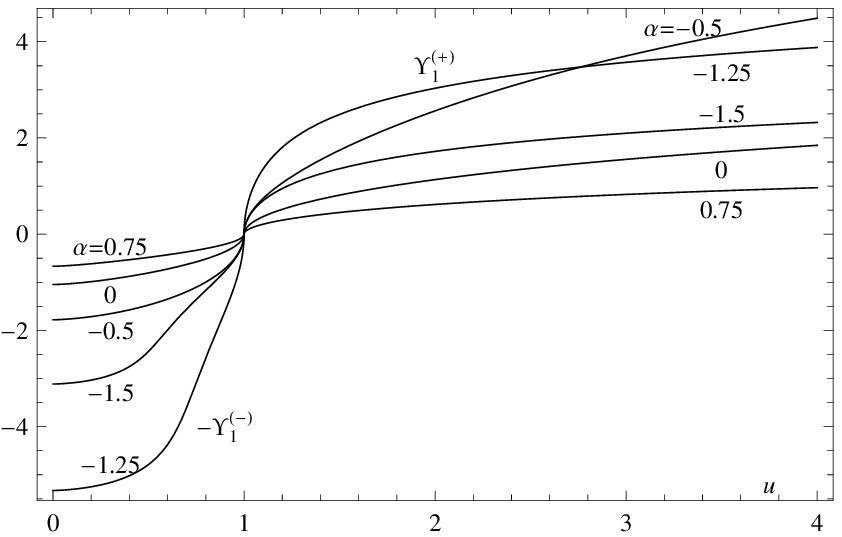}  
 \includegraphics[height=.28\textheight,width=.48\textwidth]{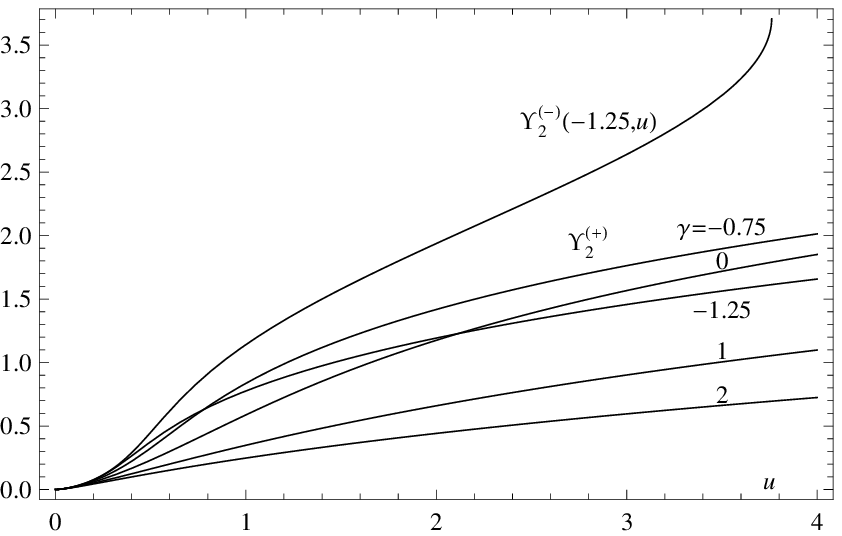}\\
 (a)\hskip 7.5cm (b)\\
 \caption{\label{FigYpsilon12} \small{Plots of the $\Upsilon$-functions: (a) $\Upsilon^{(\pm)}_{1}(\alpha,u)$;
 the $(+,-)$ correspond to $u \ge 1$, $u \le 1$ respectively.  (b) $\Upsilon^{(\pm)}_{2}(\gamma,u)$.}}
\end{figure}  

Figure \ref{FigYpsilon12} shows the typical behavior of the $\Upsilon$-functions. Note that there are only three 
independent $\Upsilon$-functions: $\Upsilon^{(-)}_{1}(\alpha,u)$ and $\Upsilon^{(-)}_{2}(\gamma,u)$ are actually the same
up to simple rescaling and translation:
\be
\Upsilon^{(-)}_{1}(\alpha,0)-\Upsilon^{(-)}_{1}(\alpha,u)=
\frac{1}{\zeta_0^{3/2}}\Upsilon^{(-)}_{2}(\frac{\alpha}{\zeta_0},\zeta_0 u)\,\, ;\,\,
\zeta_0 (\alpha)=(1+3\alpha+3\alpha^2)^{1/3}
\label{UpsminEquiv}
\ee
In the special cases $\alpha=0$ and $\gamma=0$ (which correspond to $a=0$) the $\Upsilon$-functions get the elementary 
forms:
\begin{eqnarray} \nonumber
\Upsilon^{(+)}_{1}(0,u)=\frac{2}{3}\cosh^{-1}(u^{3/2})\,\,,\,\,\,\,
\Upsilon^{(-)}_{1}(0,u)=\frac{2}{3}\cos^{-1}(u^{3/2})\\
\Upsilon^{(+)}_{2}(0,u)=\frac{2}{3}\sinh^{-1}(u^{3/2})\,\,,\,\,\,\,
\Upsilon^{(-)}_{2}(0,u)=\frac{2}{3}\sin^{-1}(u^{3/2})
\label{UpsElementary}.
\end{eqnarray} 
These correspond to the constant Ricci solutions that were mentioned in the previous section following
Eq. (\ref{ConstRicci-BG}). The fact that $a=0$ corresponds to constant Ricci solutions can be also inferred by
noticing that in this case the potential function $V_{eff}(B)$ of Eq. (\ref{EffPotB}) reduces to that in
Eq. (\ref{ConstRicci-BG}). 

Two of the three kinds of the solutions of Eq. (\ref{FirstOrderMechbeta}) can be expressed 
in terms of the $\Upsilon_{1}$ functions as
\be
\sqrt{\frac{2}{3}}\frac{|c|}{B_0^3} (r-r_0)=\Upsilon^{(\pm)}_{1}(\frac{a}{c}B_0,\frac{B}{B_0})
\label{Bsol}
\ee 
where the $\pm$ corresponds to the $\epsilon$ value. The $\epsilon=+1$ solutions are open with a minimum at $B=B_0$ 
while the $\epsilon=-1$ ones are closed (that is bounded from above) for $E<-a^3/3c$. Note that in this case $B(r)$
can be extended by ``reflection'' with respect to the $r=r_0$ line, but $L(r)=B'$ which vanishes there, reduces the 
domain of the solutions such that $B(r)$ is monotonic. We will not elaborate on this kind of solutions further.

\begin{figure}[!t]
\centering
\leavevmode\epsfxsize=10.0cm
 \includegraphics[height=.28\textheight,width=.48\textwidth]{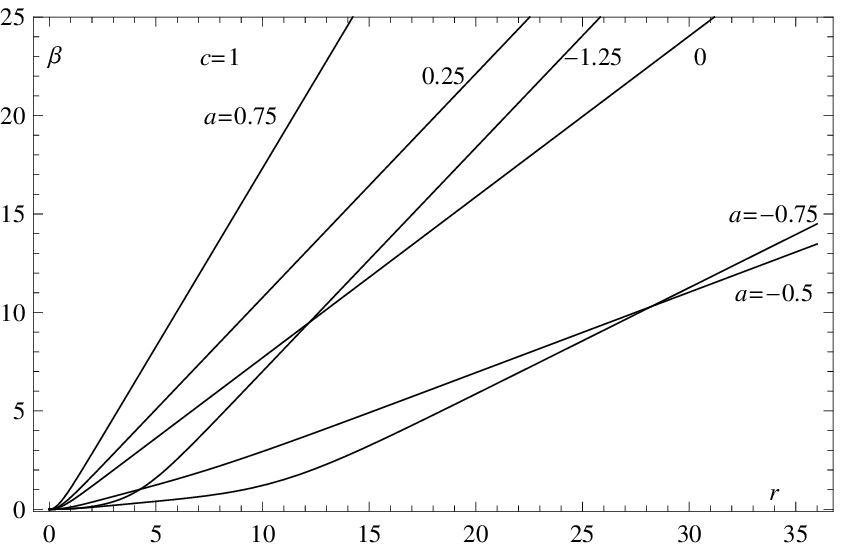}  
 \includegraphics[height=.28\textheight,width=.48\textwidth]{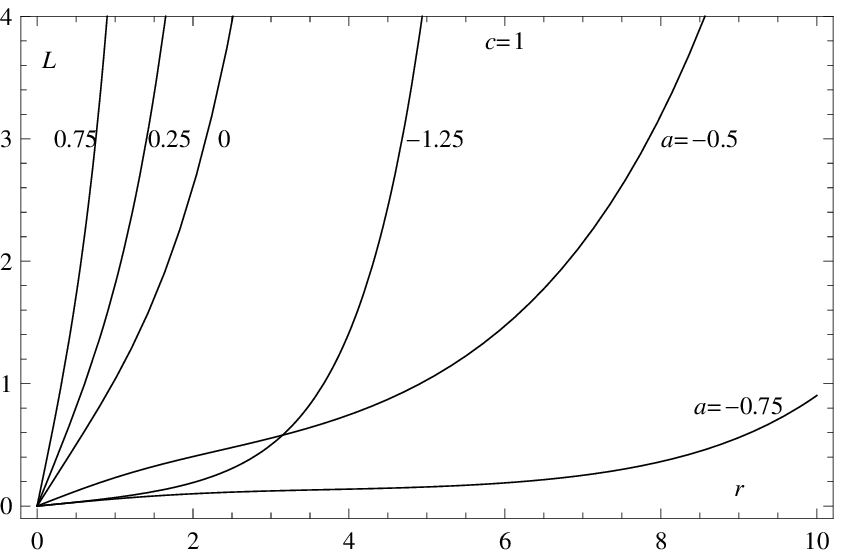}\\
 (a)\hskip 7.5cm (b)\\
 \caption{\label{FigBsols} \small{Typical solutions for $\epsilon=1$. Remember that $L=B'$. The other 
 parameters are $c=1$ and $a=-1.25,\, -0.75, \,-0.5, \, 0,\, 0.25, \, 0.75$. Note the non-monotonic dependence
 of the $\beta$-slope on $a$.}}
\end{figure}  

\begin{figure}[!b]
\centering
\leavevmode\epsfxsize=10.0cm
\includegraphics[height=.27\textheight,width=.47\textwidth]{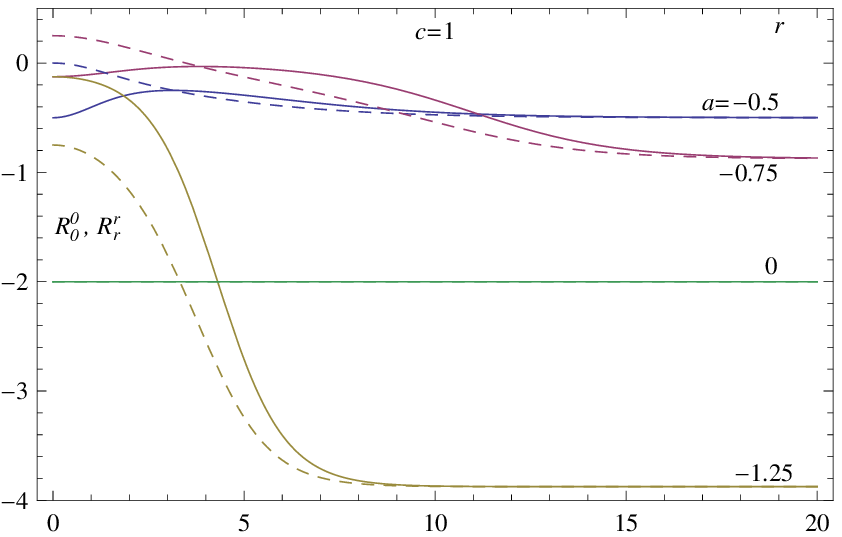}  
 \includegraphics[height=.275\textheight,width=.48\textwidth]{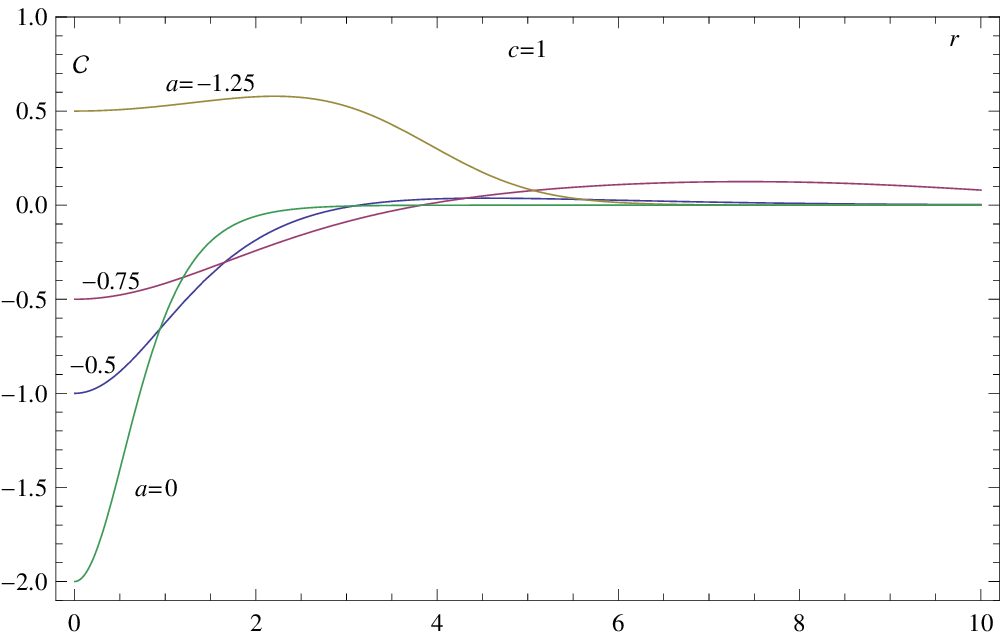}\\
 (a)\hskip 7.5cm (b)\\
\caption{\label{FigRicci} \small{(a) The Ricci components $R^0_0$ (solid) and $R^r_r$ (dashed). (b) The
Weyl tensor component ${\cal C}$. All plots for $\epsilon=1$, $c=1$ and 
$a=-1.25,\,-0.75,\,-0.5,\,0$. These are 4 of the solutions shown in Fig. \ref{FigBsols}. The curves for the 
other two cases are similar, but have much more negative Ricci components.}}
\end{figure} 

Fig. \ref{FigBsols} shows typical solutions for $\epsilon=1$ with the boundary condition $B(0)=1$ and $B'(0)=0$. The 
components of the corresponding Ricci tensor are presented in Fig. \ref{FigRicci}. They are all compatible with the 
asymptotic condition (\ref{AsymptAdS}) as should be the case. Indeed, it is easy to find from (\ref{Ricci-BG}) that 
\be
\kappa=-8c^2 \zeta^3
\label{kappa}
\ee 
where $\zeta$ is the ``energy excess'' parameter defined above. Since $\zeta^3>0$ for the open 
solutions, all of them are asymptotically ``negatively curved'' (that is $\kappa<0$). 

\begin{figure}[!b]
\centering
\leavevmode\epsfxsize=10.0cm
 \includegraphics[height=.28\textheight,width=.48\textwidth]{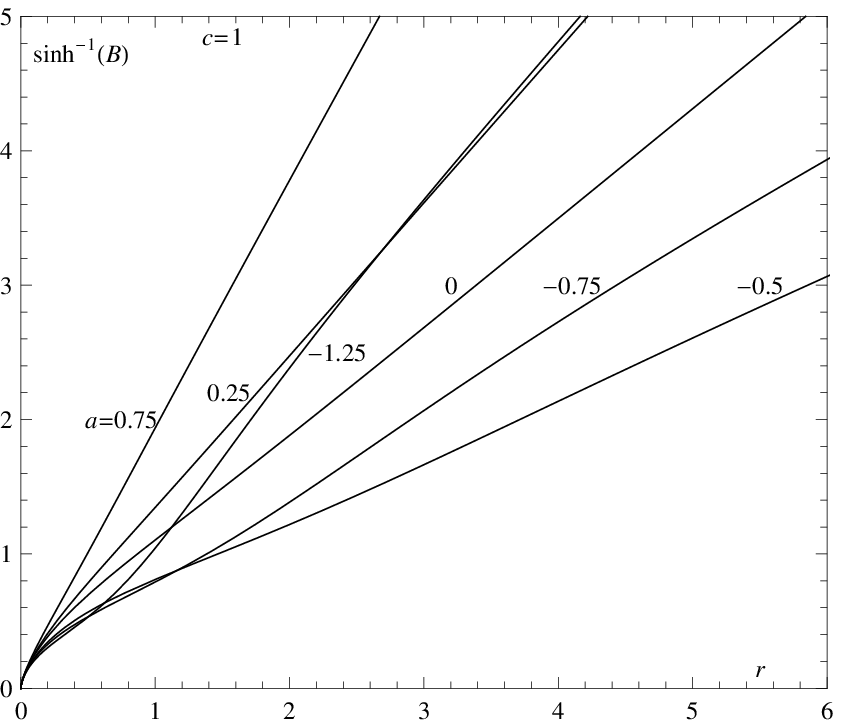}  
 \includegraphics[height=.28\textheight,width=.48\textwidth]{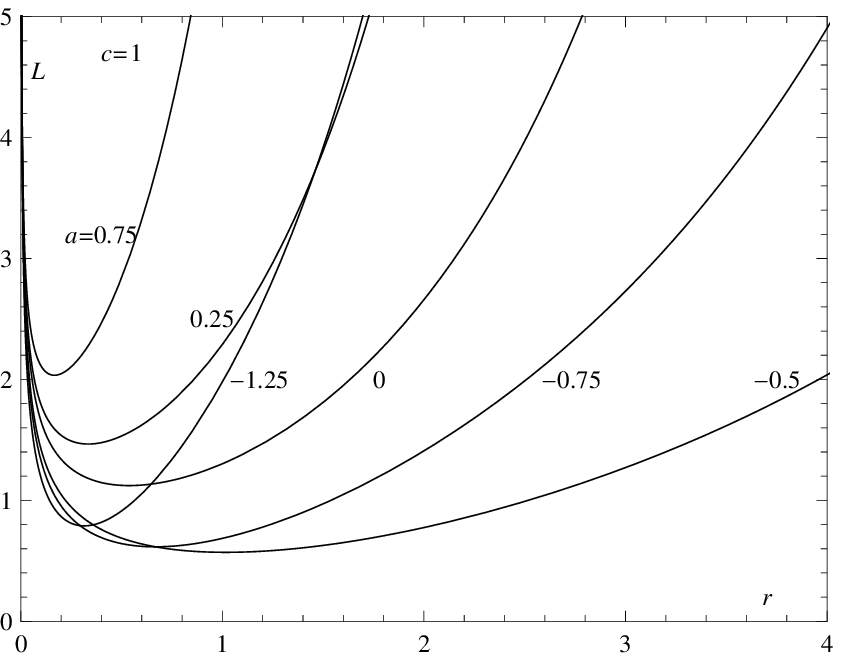}\\
 (a)\hskip 7.5cm (b)\\
 \caption{\label{FigBLminsols} \small{Typical open solutions for $\epsilon=-1$. Remember that $L=B'$. The other 
 parameters are $c=1$ and $a=-1.25,\, -0.75, \,-0.5, \, 0,\, 0.25, \, 0.75$. Note again the non-monotonic dependence
 of the slopes on $a$.}}
\end{figure} 

A special member of this family is the AdS soliton which corresponds to $a=0$ and has constant Ricci 
components, as it solves Eq. (\ref{AdSEq}) or (\ref{ConstRicci-BG}). Note however, that unlike (A)dS space, 
it is not conformally flat, as is also obvious from the corresponding curve in Fig. \ref{FigRicci}b.
 
The third type of solutions is with $\epsilon=-1$ and $E\ge-a^3/3c$, so they can have the whole range $0\le B<\infty$. 
It may be written simply using the energy excess parameter $\zeta$ which is positive in this case too:
\be
\sqrt{\frac{2}{3}}|c|\zeta^{3/2} (r-r_0)=\Upsilon^{(+)}_{2}(\frac{a}{c\zeta},\zeta B)
\label{Bsol2}
\ee 

\begin{figure}[!t]
\centering
\leavevmode\epsfxsize=10.0cm
\includegraphics[height=.275\textheight,width=.51\textwidth]{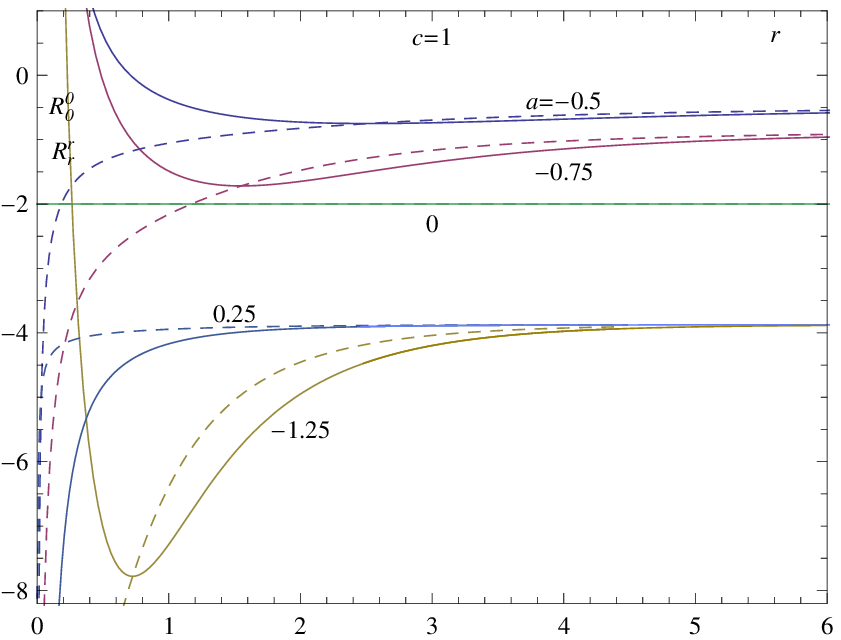}  
 \includegraphics[height=.275\textheight,width=.45\textwidth]{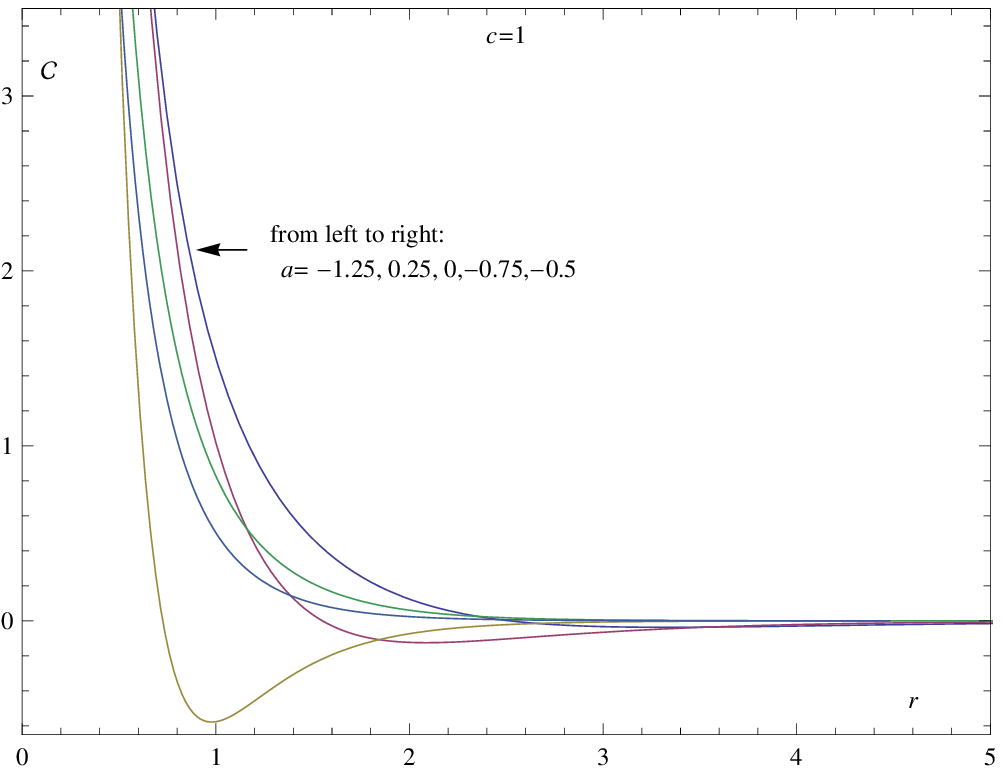}\\
 (a)\hskip 7.5cm (b)\\
\caption{\label{Fig2Ricci} \small{(a) The Ricci components $R^0_0$ (solid) and $R^r_r$ (dashed). (b) The
Weyl tensor component ${\cal C}$. All plots for $\epsilon=-1$, 
$c=1$ and $a=-1.25,\, -0.75,\,-0.5,\,0,\,0.25$. These are 5 of the solutions shown in Fig. \ref{FigBLminsols}. 
The $a=0.75$ curves  are similar to the $a=0.25$ ones, but have much more negative Ricci components.}}
\end{figure}

Fig. \ref{FigBLminsols}
shows typical solutions of this kind where we choose $B(0)=0$. Fig. \ref{Fig2Ricci} has the components of the corresponding 
Ricci tensor which shows a singularity on the axis. However, these solutions  may still be physically relevant as exterior 
solutions of appropriate cylindrically-symmetric sources. We took the same values as in Fig. \ref{FigBsols} for the
parameters $a$ and $c$ as well as for the ``energy excess'' $\zeta$. We also chose to present $\sinh^{-1}(B)$ 
instead of $B$ in order to cover both the small $B$ and large $B$ regions. The exponential increase of 
$B(r)$ away from the axis is obvious. An exception from the generic exponential behavior is the minimal 
$E$-value for unbounded solutions which correspond to $\zeta=0$. In this case the asymptotic behavior is $B \sim r^2$.
The asymptotic behavior of the Ricci components is given again by (\ref{AsymptAdS}) with (\ref{kappa}) so in both cases
$\kappa<0$. The $\zeta=0$ solution is asymptotically Ricci flat.

Note that again the special case $a=0$ allows for explicit (constant Ricci) solutions in terms of elementary 
functions - see (\ref{UpsElementary}).
For the ``last unbounded trajectory'', which corresponds to  $\zeta=0$ (with $\epsilon=-1$) 
we find $B=[\sqrt{3/2}|c|(r-r_0)]^{2/3}$. This solution is not only asymptotically Ricci flat, but has $R_{\mu\nu}=0$ 
identically and is actually the well-known ``Kasner'' string-like solution of ordinary GR.
Another special case which requires anyhow a separate treatment is $c=0$. In this case we can just integrate Eq. (\ref{r(B)})
directly to get the general solutions 
\begin{eqnarray} 
B=B_0\cosh^2\left(\frac{|a|(r-r_0)}{\sqrt{2B_0}}\right) \,\,,\,\,\,\,\epsilon=1 \\ \nonumber
B=B_0\cos^2\left(\frac{|a|(r-r_0)}{\sqrt{2B_0}}\right) \,\,,\,\,\,\,\epsilon=-1 \,\,,\,\,\,\, E<0
\label{BsolBounded}
\end{eqnarray}
instead of (\ref{Bsol}), while instead of (\ref{Bsol2}) we have
\be
B=\frac{a^2}{E}\sinh^2\left(\sqrt{E/2} \,\,\,\, (r-r_0)\right) \,\,,\,\,\,\,\epsilon=-1 \,\,,\,\,\,\, E> 0  
\label{BsolUnBounded}
\ee
The special case $E=0$ can be obtained from the limit of this equation to give a  behavior of
\be
B= a^2 (r-r_0)^2 /2  \,\,,\,\,\,\,\epsilon=-1 \,\,,\,\,\,\, E= 0  
\label{BsolUnBoundedE0}
\ee
This family is supplemented by the exponential solution (\ref{SolExp}) encountered already at the beginning.

All these $c=0$ solutions have non-trivial curvatures, but as mentioned above they are conformally flat. However,
they are not trivial even within the framework of CG, since they are either conformal to Minkowski space-time with
a conic angular deficit like (\ref{BsolUnBoundedE0}), or to a certain region of it like the others. The solution 
(\ref{BsolUnBoundedE0}) is therefore the analog in this gauge of CG to the conical string-like solution of GR.

Finally we note a possible generalization of this work to the scalar-tensor extension of CG which allows for a 
breakdown of the conformal symmetry. Some cylindrically-symmetric solutions of this kind were obtained numerically
in a recent study \cite{BrihayeVerbinCyl}, but considering
the results presented in this section, it may turn out possible to obtain explicit exact solutions in the gauge where the 
scalar field is constant. The metric tensor will have now two independent components, but the second may be taken as a 
conformal factor (as in ref \cite{BarabashSht1999}) which will simplify the equations so that analytical treatment will 
be possible.

\section{Null Geodesics} \label{geod}
\setcounter{equation}{0}

We proceed here to examine further the nature of these solutions by studying their geodesics. The 
geodesic equations in the metric $diag(B^2, -1, -L^2, -B^2)$ are easily integrated to give
\begin{eqnarray} 
\dot{z}=k\,\,,\,\,\,\,
L^2\dot{\varphi}/B^2=\ell \,\,,\,\,\,\,
\dot{r}^2+(k^2-1)B^2+\ell^2 B^4/L^2=-\mu B^4/{\cal E}^2
\label{geodesics1}
\end{eqnarray} 
where the coordinates are functions of time $t$ and $k<1$, $\ell$  and ${\cal E}$ are constants of the motion. 
The parameter $\mu=0,1$ distinguishes between lightlike or timelike geodesics respectively. Since only
null geodesics have an invariant meaning, we will concentrate on them.

One can analyze numerically the trajectories by the effective potential for the $r$-motion which satisfies
$\dot{r}^2/2+U_{eff}(r)=0$. However, analytic treatment 
is possible if we ``compromise'' on a parametric representation of the solutions $r(t)$ by $(r(B),t(B))$. We will not 
expect any trouble with that since we concentrate in the open solutions where $B(r)$ is monotonic. The $r$-equation
will be replaced by $\dot{B}^2/2+W_{eff}(B)=0$ with
\be
W_{eff}(B)=\left[\frac{\ell^2}{2}+(1-k^2)\left(\frac{\epsilon c^2 \alpha^3}{3}-E\right)\right]B^4+
\epsilon c^2(1-k^2)\left(\alpha^2 B^2+\alpha B + \frac{1}{3}\right)B
\label{Nullgeodesics}
\ee
where $\alpha=a/c$. This is the potential for the general case with $c\ne 0$. The $c=0$ case needs as usual a special 
treatment and is given by (\ref{Nullgeodesicsc0}) below. 
Actually, since there exist explicit solutions in this case one may obtain also $U_{eff}(r)$ explicitly.
It is straightforward to analyze the different trajectory types from $W_{eff}(B)$ and especially its zeroes. 
The solutions may be bounded (in $B$ or in $r$) where  $W_{eff}(B)<0$ in a finite interval or 
unbounded (or open) if this interval extends to infinity. The process is simplified by noticing that $W_{eff}(B)$ 
has only one zero in addition to the possible one at $B=0$.

\begin{figure}[!b]
\centering
\leavevmode\epsfxsize=10.0cm
 \includegraphics[height=.28\textheight,width=.48\textwidth]{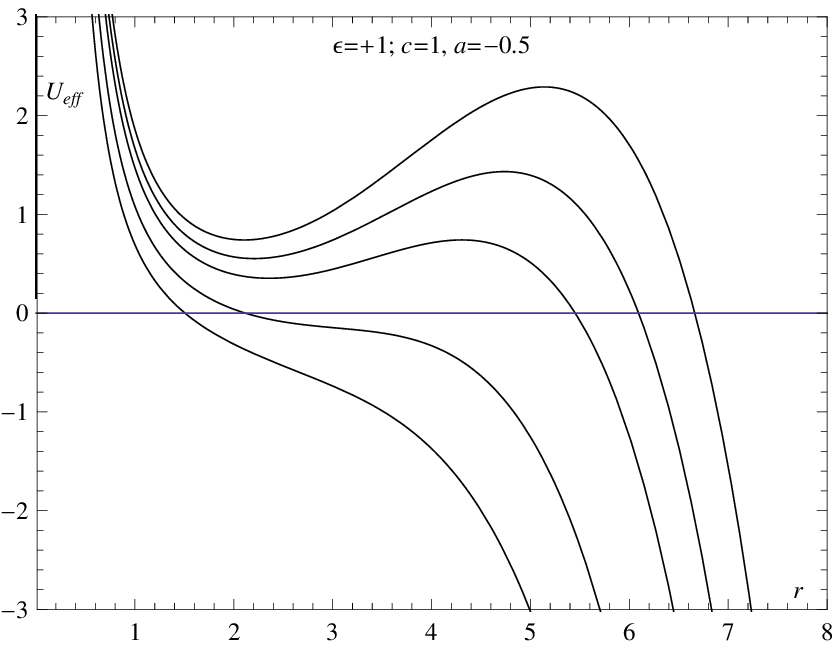}  
 \includegraphics[height=.28\textheight,width=.48\textwidth]{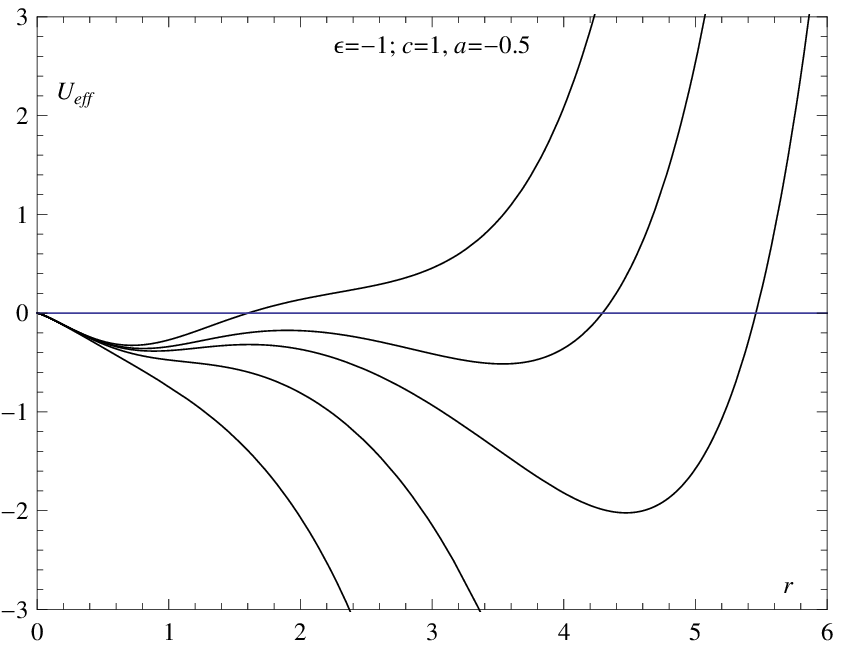}\\
 (a)\hskip 7.5cm (b)\\
 \caption{\label{FigBUeff} \small{Effective potential (divided by $(1-k^2)/2$) curves for null geodesics in the two types 
 of metrics with $c=1$ and $a=-0.5$. 
 (a)  $\epsilon= 1$, $\xi=-0.15,\;  -0.13,\;  -0.11, \; -0.10,\;  -0.09$ ;
 (b) $\epsilon= -1$, $\xi=-0.2,\; 0,\;  0.07,\;  0.1,\;  0.15$.
 In both cases the higher curves correspond to the larger values of $\xi$.}}
\end{figure} 

For $\epsilon=1$ all solutions are open, namely $B_{min} \le B < \infty$ where $B_{min}$
depends on the constants of the motion $k$ and $\ell$ and on the other geometrical parameters $a$, $c$ and $E$ 
(see below). A necessary condition for solutions to exist (for $\epsilon=1$) is that the coefficient of the $B^4$ 
term in $W_{eff}(B)$ which is proportional (with the positive coefficient $c^2(1-k^2)$) to
\be
\xi=\frac{3\ell^2}{2c^2(1-k^2)} -\frac{3E}{c^2}+\epsilon \alpha^3 
\label{xi}
\ee 
will be negative, otherwise there will be no zeroes of $W_{eff}(B)$ for $B \ge 1$ which is the domain of $B$ in this case. 
This imposes an upper bound on the combination $\ell^2/(1-k^2)$:
\be
\frac{\ell^2}{1-k^2} < 2\left( E-\frac{c^2 \alpha^3}{3} \right)
\label{UpperBound}
\ee 
For $\epsilon=-1$ there are two cases: open solutions which now extend for $0 \le B < \infty$ occur if 
Eq. (\ref{UpperBound}) is satisfied. In the complementary case we have bounded solutions for $0 \le B < B_{max}$
where $B_{max}$ depends on the characteristic parameters. $B_{max}$ and also $B_{min}$ can be written together as
\be
B_{min/max}=
\frac{\alpha^2}{\epsilon\xi} \left( -1+ \sqrt[3]{\epsilon\xi/\alpha^3 -1}- \sqrt[3]{(\epsilon\xi/\alpha^3-1)^2} \right)
\label{Bminmax}
\ee 
with the corresponding value of $\epsilon$. 

\begin{figure}[!b]
\centering
\leavevmode\epsfxsize=10.0cm
 \includegraphics[height=.28\textheight,width=.40\textwidth]{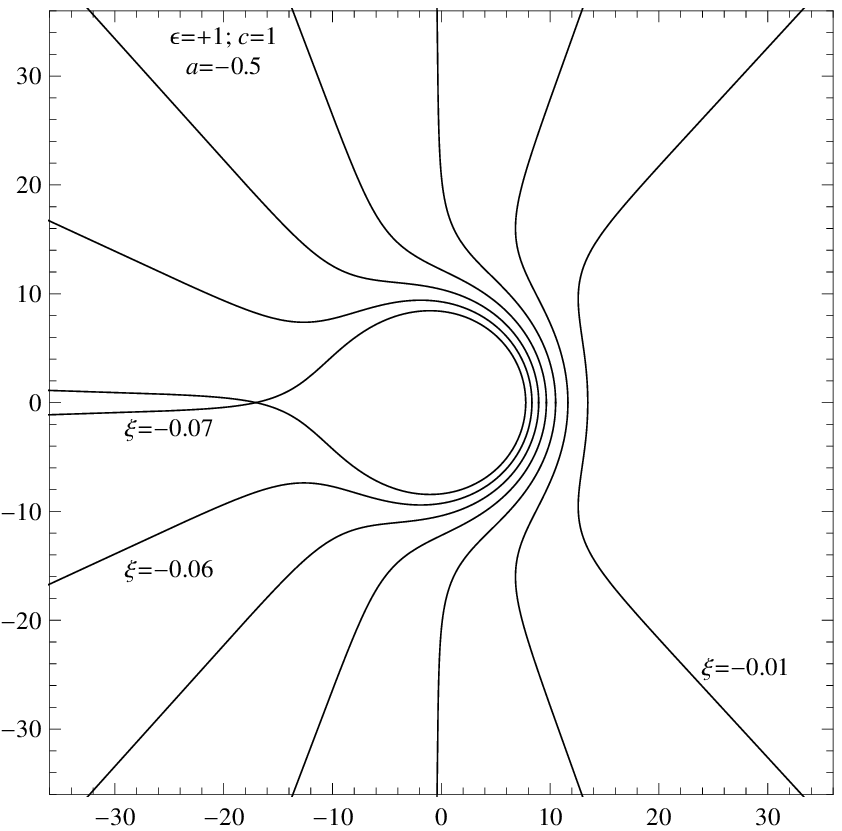} \hspace{1cm} 
 \includegraphics[height=.28\textheight,width=.40\textwidth]{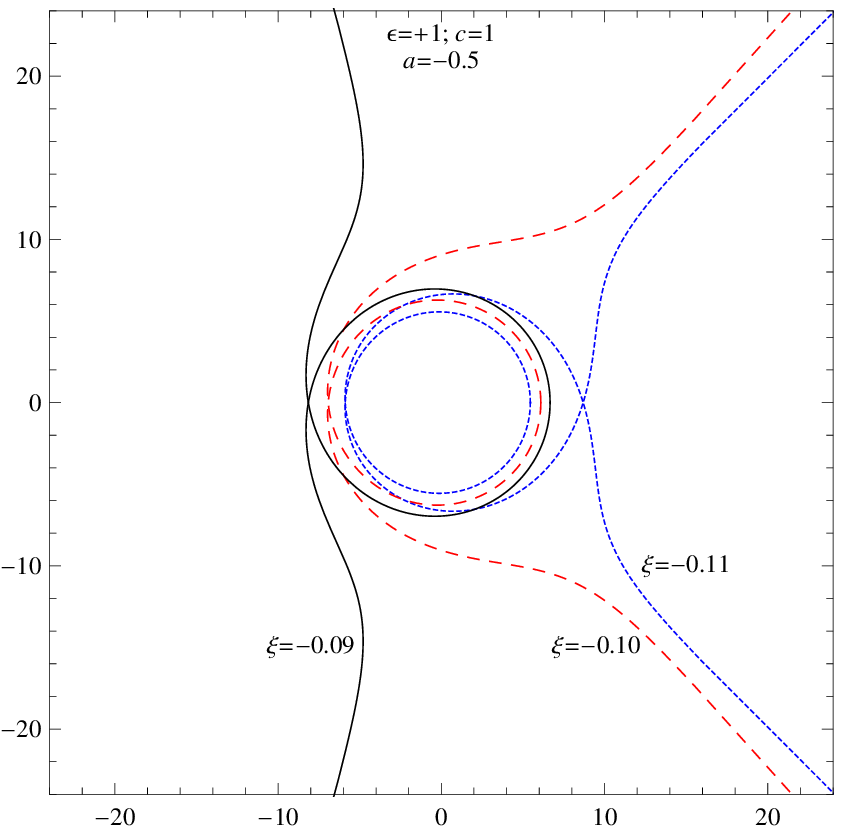}\\
 (a)\hskip 7.5cm (b)\\
 \caption{\label{FigEpsPLTraj} \small{Projection in the $(r,\varphi)$ plane of null trajectories with  
 $\epsilon=+1$, $c=1, \;a=-0.5$.  (a) Solutions with $\xi$-values from $\xi=-0.07$ to $\xi=-0.01$ in even steps;
 (b) Three more winding solutions with $\xi=-0.11,\;-0.10,\;-0.09$.}}
\end{figure}

Plots of $U_{eff}(r)$ which present all these properties are shown in Fig \ref{FigBUeff}. For $\epsilon= 1$
only $\xi<0$ curves are shown since otherwise $U_{eff}(r)$ is always positive and no solutions exist. Note that each
curve in this figure corresponds to a single null geodesic (up to $t$-translation) defined by $k$ and $\ell$ since 
only vanishing ``effective energy'' is allowed.

In order to find out about the shape of the geodesics, one may solve for $r(\varphi)$ or equivalently for the parametric 
representation $(r(B),\varphi(B))$. It turns out to be simpler to transform to $v=1/B$ where we find the following first 
order equation for light-like trajectories, or more accurately, their projection on the $(r,\varphi)$ plane:
\be
\frac{1}{2}\left(\frac{dv}{d\varphi}\right)^2+
2\left[1-\frac{2(1-k^2)}{\ell^2}\left(E-\frac{\epsilon c^2 }{3} (\alpha+v)^3\right)\right]
\left(E-\frac{\epsilon c^2 }{3} (\alpha+v)^3\right)^2 = 0
\label{lightlike trajectories}
\ee

\begin{figure}[!t]
\centering
\leavevmode\epsfxsize=10.0cm
 \includegraphics[height=.28\textheight,width=.40\textwidth]{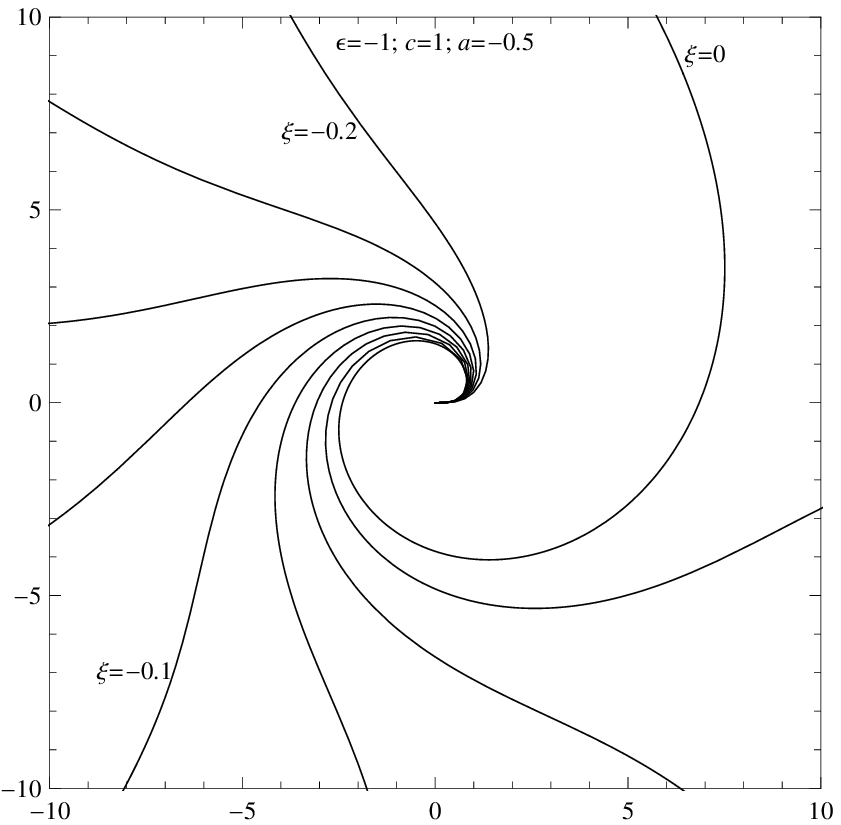}  
 \includegraphics[height=.28\textheight,width=.48\textwidth]{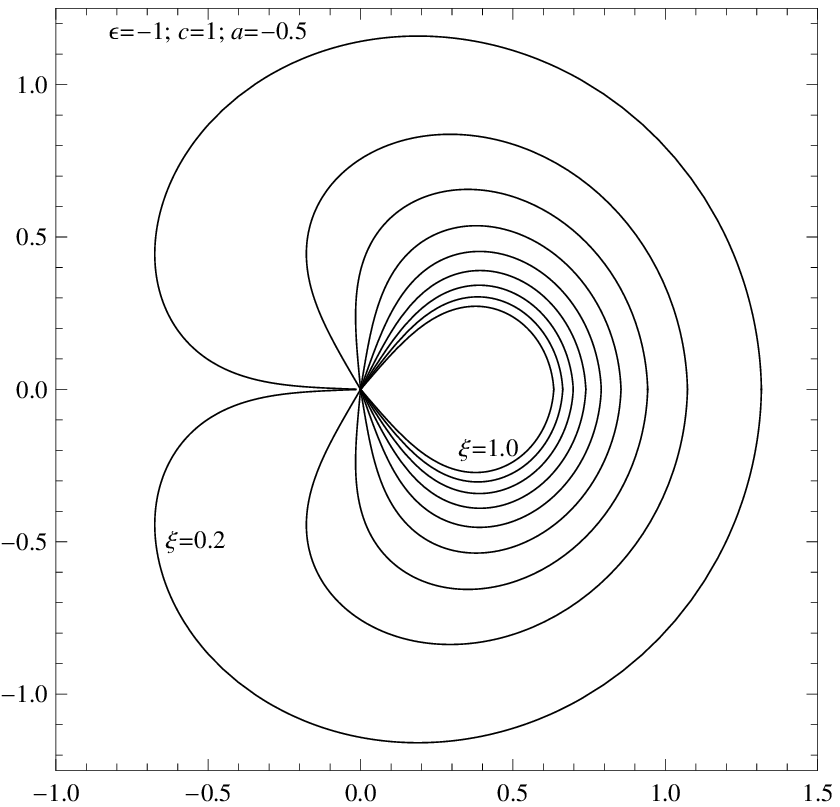}\\
 (a)\hskip 7.5cm (b)\\
 \caption{\label{FigEpsMinTraj} \small{Projection in the $(r,\varphi)$ plane of null trajectories with  
 $\epsilon=-1$ , $c=1, \;a=-0.5$.  (a) Open solutions with $\xi$-values from $\xi=-0.2$ to $\xi=0$ in even steps;
 (b) Bounded solutions with $\xi$-values from $\xi=0.2$ to $\xi=1$ in even steps.}}
\end{figure}

The projections of some typical trajectories appear in Figs. \ref{FigEpsPLTraj}-\ref{FigEpsMinTraj}. 
They demonstrate explicitly the features obtained from the
general discussion above. The $\epsilon=1$ solutions are all open with a ``periastron'' chosen to be always at $\varphi=0$.  
Several windings are possible according to the values of the parameters. The $\epsilon=-1$ solutions may be either open with
$0\le r <\infty$, or bounded with $0\le r \le r_{max}$. Again, several windings are possible.

Finally we discuss shortly some special cases. The first is $a=0$. Generally null geodesics in this case
are just slightly different from the 
adjacent solutions except when also $\xi=0$ for $\epsilon=-1$. This is the only case when $U_{eff}(r)$ vanishes 
asymptotically to allow open geodesics with $\dot{r}(t)\rightarrow 0$ as $t \rightarrow \infty$. 
The resulting orbits are spirals which become denser  
as $r$ increases. Actually, the effective potential $U_{eff}(r)$ can be written explicitly in terms of elementary
functions, and the typical null geodesics with $r(0)=0$ and $\varphi(0)=0$ are 
\begin{eqnarray} \nonumber
r(t)=\sqrt{\frac{2}{3}}\frac{1}{c}\sinh^{-1}(\bar{t}^3) \,\,,\,\,\,\,
\varphi(t)= \frac{3\ell}{2c^3}\sqrt{\frac{6}{1-k^2}}\left[\bar{t}-\frac{1}{6}\left(\tan^{-1} 
(2\bar{t}-\sqrt{3})+\tan^{-1} (2\bar{t}+\sqrt{3})\right)  \right.  \\  \left.  -\frac{1}{3}\tan^{-1} \bar{t}   
+ \sqrt{3}\log \left( \frac{\bar{t}^2-\sqrt{3}\bar{t}+1}{\bar{t}^2+\sqrt{3}\bar{t}+1} \right)\right]
\label{a0trajectory}
\end{eqnarray}
where $\bar{t}=\sqrt{(1-k^2)/6}\; ct$.

The case $c=0$ also allows to obtain explicit expressions for $U_{eff}(r)$, but it is easier to solve for $B(t)$
using the effective potential of Eq. (\ref{Nullgeodesics}) which now simplifies to
\be
W_{eff}^{^{(c=0)}}(B)=\left(\frac{\ell^2}{2}-(1-k^2)E\right)B^4+ \epsilon (1-k^2)a^2 B^3
\label{Nullgeodesicsc0}
\ee
The transformation $v=1/B$ gives a mechanical equation with a linear potential which gives easily the following 
solutions in the various cases. Two of them can be written as 
\begin{eqnarray} 
\frac{1}{B}=-\frac{\epsilon(1-k^2)a^2 t^2}{2} +\frac{\epsilon}{a^2}\left(E-\frac{\ell^2}{2(1-k^2)}\right)
\label{c0trajectoryBounded}
\end{eqnarray}
where we chose $t=0$ to be the time which corresponds to the minimal or maximal radial distances for $\epsilon=1$
or $\epsilon=-1$ respectively. For $\epsilon=1$, $t$ should be further restricted such that $1/B$ is non-negative.
 Note that the relative sizes of the two quantities in the bracket of the second term
are such that it is always positive. The third kind of geodesics which have $0<B<\infty$ exist only for $\epsilon=-1$
and are given by 
\begin{eqnarray} 
\frac{1}{B}=\frac{(1-k^2)a^2 t^2}{2} + \sqrt{2(1-k^2)E-\ell^2}\; t
\label{c0trajectoryUnBounded}
\end{eqnarray}
In order to get the explicit $r(t)$ dependence of these geodesics one needs the corresponding $B(r)$ solutions for this 
case - Eqs. (\ref{BsolBounded})--(\ref{BsolUnBounded}). We will not present here this last step. 
The explicit solution $\varphi(t)$ which we skip too may be obtained by direct integration
of (see (\ref{geodesics1}))
\be
\frac{d\varphi}{dt}=\frac{\ell B^2}{L^2}=\frac{\ell}{2(E-\epsilon a^2/B)}
\ee
with $1/B(t)$ obtained from (\ref{c0trajectoryBounded})--(\ref{c0trajectoryUnBounded}) above.

\section{Conclusion} \label{Concl}
\setcounter{equation}{0}

We have analyzed the vacuum field equations of CG in the static cylindrically-symmetric case and found all solutions 
explicitly. In some cases the solutions are expressed in terms of elementary functions.  

There are three kinds of solutions: two open ones and one closed. The open ones split into a regular family 
(with a possible conic singularity on the axis) which contains the 
AdS soliton and a family of singular solutions where $g_{00}$ vanishes on the axis and the Ricci and Weyl components 
diverge there. These two families of open solutions have generically asymptotically vanishing Weyl tensor. 
Some special cases are conformal to conical space-times.

We have further analyzed the null geodesics in the two families of the open spacetimes. We found that the singular
spacetimes support two kinds of null geodesics: open ones (that is $0<r<\infty$) which are usually spirals when the angular 
momentum is below a certain maximal value, and bounded ones when the angular momentum is above the critical value.
In the regular spacetimes only open geodesics exist for the same range of angular momentum. Unless the angular momentum
vanishes, the orbits in this case avoid $r=0$. Above the critical value no geodesics exist at all. 

A possible future application of these results is to use these families of exact solutions to analyze light bending in the 
vicinity of localized linear sources in CG. The analogous problem of light bending in spherically-symmetric gravitational
 field was only recently settled \cite{Sultana+Kazanas}.

\end{document}